# Reflections on organization, emergence, and control in sociotechnical systems

Vincenzo De Florio

MOSAIC research group
iMinds research institute & University of Antwerp
Middelheimlaan 1, 2020 Antwerp, Belgium
e-mail: vincenzo.deflorio@gmail.com

## Abstract

Human and artificial organizations may be described as networks of interacting parts. Those parts exchange data and control information and, as a result of these interactions, organizations produce emergent behaviors and purposes — traits the characterize "the whole" as "greater than the sum of its parts". In this chapter it is argued that, rather than a static and immutable property, emergence should be interpreted as the result of dynamic interactions between forces of opposite sign: centripetal (positive) forces strengthening emergence by consolidating the whole and centrifugal (negative) forces that weaken the social persona and as such are detrimental to emergence. The result of this interaction is called in this chapter as "quality of emergence". This problem is discussed in the context of a particular class of organizations: conventional hierarchies. We highlight how traditional designs produce behaviors that may severely impact the quality of emergence. Finally we discuss a particular class of organizations that do not suffer from the limitations typical of strict hierarchies and result in greater quality of emergence. In some case, however, these enhancements are counterweighted by a reduced degree of controllability and verifiability.

# 1 Introduction

A well-known statement attributed to Aristotle is "the Whole is greater than the sum of its Parts" (Aristotle, 2014). This concise statement hints at one of the grand principles of nature: **Emergence**, or the natural development of complex traits, patterns, or behaviors, from the organization and interactions of a set of simpler components. Emergence refers to the manifestation of new properties that are often not predictable from an analysis of those lower-level components. Emergence is a phenomenon taking place at all natural scales. A classic example is the emergence of the liquid state from a combination of hydrogenous and oxygen. A second example is given by considering the human being: A Whole whose sophisticated traits set her apart from all other beings. The human being is the result of an organization of simpler Parts — the body organs and systems — each of which is in turn a Whole greater than the sum of its own constituents.

So great is the amount of available examples of "substances" (De Florio, 2014a) for which the celebrated Aristotelian statement applies that one could be tempted to assert its unconditional validity — were it not for the many counterexamples that are also available. Let us consider for instance the case of human organizations. Human history and everyday's practice provide us with many examples of sub-optimal organizations in which the quality of the Whole often degrades to the point that the system ceases to make sense and disintegrates into its Parts. Examples of this may be found, e.g., among political systems, business bodies, and disaster management organizations (De Florio, Sun, & Blondia, 2014). Because of this, rather than "the sum of its Parts," it may make more sense to say that the Whole is the dynamic *product* resulting from the *interaction* of constituent parts arranged according to an *organizational structure*, and that this product may be characterized by different *qualities of emergence*.

In other words, we argue that the Whole may be interpreted as the result of a dynamic process of interacting "agents" — namely Parts characterized by various degree of behavior — ranging from purposeful predefined behavior to proactive autonomic behavior — that, depending on several factors may either contribute or counteract towards emergence. Factors playing a role in this process include, among others: Consciousness, cybernetic level, individual agenda, system of priorities, system of beliefs, cultural aspects, system-specific aspects (e.g. anthropological factors), wear-out (aging) factors, organization (e.g. hierarchies, heterarchies, and communities), the match with the cybernetic level of the organization, the presence of "win-win" conditions (e.g., mutualistic or commensalistic relationships, such as symbiosis), and the various environmental factors. Quality of emergence (QoE) represents a measure of the *resultant force* characterizing the above-mentioned dynamic process. Said resultant force is interpreted in what follows as the algebraic sum of opposite forces (Dominici, 2011, 2013, 2014):

- Centripetal forces, leading to strengthening or evolving the Whole and to an increase in the QoE.

- Centrifugal forces, "weakening" the "social persona" and the identity of the Whole and detrimental to its QoE.

Obviously, QoE is a generic term that may translate into different desireable behaviors and properties including, among others: High performance, low energy consumption, extended operating time, advanced teleological behaviors, welfare, resilience, autonomicity, survivability, determinism, and controllability. Our focus in what follows will be that of QoE as a measure of controllability, resilience, and cybernetic behaviors. Particular attention will be reserved to the roles played by the organization, the individual behaviors, and the environmental conditions, towards the emergence of a "greater Whole."

This chapter is structured as follows: Section 2, recalling the lesson of Leibniz, highlights the major intrinsic and extrinsic factors that are responsible for the dynamic evolution of QoE. In Sect. 3 the attention is on hierarchies and their persona, viz. the embodiment (or personification) of their Wholes. In this section examples in social and biological systems are also used to highlight a number of extrinsic factors corresponding to high or low QoE. A major extrinsic factor negatively affecting QoE is then introduced and two dualistic strategies to improve QoE are identified. One such strategy is exemplified in the two organizations surveyed in Sect. 4: sociocracy and fractal social organizations. Section 5 finally concludes with a summary of our reflections.

## 2  Quality-of-Emergence and Its Causes

As observed by Leibniz (Leibniz & Strickland, 2006), the emerging quality of the Whole depends on both structural and contingent causes:

- Structural causes reflect the systemic properties of the Whole: Its organizational, behavioral, and architectural traits. This is the "abstract model" of the Whole, so to say, which corresponds to what Leibniz referred to as *Monads* — indivisible metaphysical conceptual units (De Florio, 2014a).

- Contingent causes come into play when a Monad is "implemented", namely when it is "*real*ized" in terms of physical parts. This implementation or materialization reveals an organization of constituent parts (Deleuze, 1993). Contingent causes then represent the match between the physical implementation of the Whole (between one of its possible "codifications" or "materializations") and *the context*, namely the dynamic variation in the conditions of a deployment environment.

A special class of contingent causes is given by what Leibniz referred to as "compossibility", namely the possibility for coexistence of multiple instances of different Monads. A classic example of this mutual compatibility is given by trophic interactions (Temkin & Eldredge, 2014) such as those between predator and prey species.

One of the many merits of Leibniz was that of devising an *economic model of coexistence* among monads and their physical realizations. In facts, Monads are "selected for existence" in accordance to their QoE. This selection pressure operates within a regime of competition: The "world" can only host a finite number of Monads[1], and "God" chooses which Monads to persist on the base of their structural (intrinsic) and contingent (extrinsic) characteristics. In other words, in Leibniz, QoE becomes the evaluation criterion for being selected for existence. With modern terminology we could say that QoE determines a system-of-system's *resilience* — in other words, QoE determines the persistence of the Whole and its Parts when facing changes (Laprie, 2008).

## 3  Hierarchy, Control, and Collective Persona

We should turn our attention next to hierarchies as a specific class of organization. Hierarchies have been extensively studied by many Authors — for instance (Koestler, 1967; Temkin & Eldredge, 2014); extensive bibliographies are available, e.g., in (Walonick, 1993; Puranam & Goetting, 2009). Here emphasis is put on the intrinsic and extrinsic factors resulting in QoE and emergence failures in hierarchies.

In a strict hierarchy, the emerging "persona" (in the sense expressed in (De Florio, 2014b)) is the "root" of the hierarchy tree. A good example of this can be found in military organizations where the whole organization is "personified" in the chief or general from which all orders "descend". The behavior of the organization is in this case — ideally — the same as the behavior of the chief or general. The latter has perfect "remote control" over the latter, and the Parts are an organizational extension, or prosthesis (sensu McLuhan (Macdougall, 2013)) of the entity in control. QoE translates in the following emerging properties: Controllability and determinism. "Obedience" is the domain-specific term corresponding to the centripetal mechanism adopted by the Whole to safeguard and enforce its own QoE. Other centripetal factors are specific contextual conditions. For instance, in times of war, "priorities" might include the safety of the dear ones, the family, the city, the Country etc. Those factors may lead to a strengthening of the Whole and a greater degree of QoE. Centrifugal forces are exemplified by a mismatch between the systemic classes[2] of the Parts and that of the Whole, or by demoting the Parts to roles typical of lesser beings[3].

A different example is given by living beings endowed with at least minimal degrees of consciousness — with a "mind" that is. In this case the brain is the "leading substance" that personifies the whole body. A common vernacular is in fact "we are our brains". QoE translates again in controllability and in the "good functioning" of the "network of Parts". Centripetal and centrifugal factors are those typical of biological beings: Physical state (healthy vs. unhealthy); aging (young vs. old); environmental conditions (favorable vs. adverse); and so on.

---

1 We actually don't know whether this limitation is real or it is an artifice to steer evolvability to ever greater levels—or in other words, to let the QoE of the First Monad—-the Divine Unity—to ever increase.

2 By systemic class it is meant in this chapter the equivalence class a system belongs to in a General Systems Theory classification such as the one introduced in (Rosenblueth, Wiener, & Bigelow, 1943) or the one in (Boulding, 1956). As an example, Thermostat, Cell, and Animal are three of Boulding's systemic classes.

3 The inhuman conditions and treatment of the Italian troops in the Dolomite fields during World War I constitute dreadful examples of the lengths the armies of the past went in order to enforce their organizational rules and thus their QoE.

A third example is given by business bodies — for instance, a company — and the executive and managing director. This director is often the personification of the "root level" of the company: The board of directors, namely the planning and executive "organ" of the company. Desired emerging properties in this case include effective "remote control" throughout the hierarchy, performance, production throughput, quality, reduction of costs, and resilience, interpreted here as the ability to minimize threats and identify opportunities in a turbulent and competitive business environment. Centripetal forces are those that strengthen the "sense of belonging" and of "ownership" — stimulated, e.g., by allowing the Parts to become official stakeholders of the company. Participation to social decisions is also an important positive force (Sun, De Florio, Gui, & Blondia, 2007) as it also is whatever factor enforcing the conception that "a healthy company means a healthy employee". Such "win-win" factors contribute to the emergence of an *ecosystem of Parts* that enhances the resilience of the Whole. Centrifugal forces are basically those that lead to the opposite conditions — in particular "win-lose" ("healthy factory does not mean a healthy employee") and mismatches between Parts/Whole systemic classes.

It is important to emphasize the point that the very structure of the organization sets the boundaries and conditions[4] for the expression of centrifugal and centripetal forces. As an example, in a strict hierarchy, every level of the hierarchy is intrinsically a potential single-point-of-failure and a single-point-of-congestion for the control and data flows that constitute the functioning of the Whole (Astley & Fombrun, 1983).

Having defined a framework for the discussion of Emergence and, within said framework, its two major beneficial and detrimental factors, it is now possible to express the enhancement of QoE through the enaction of two dual classes of processes — those processes that aim at promoting win-win factors and those reducing win-lose factors. In other words, *dampening centrifugal forces* and *amplifying centripetal forces*.

Social role assignment may influence both forces. Parts may be assigned roles corresponding to behaviors that are too "elementary" with respect to their natural (i.e., systemic) capabilities (Rosenblueth et al., 1943, Boulding, 1956). In practice, this translates into demoting Parts to inferior systemic classes. Treated as "cogs within a greater cog", systemically more advanced Parts are likely to reassess the benefits deriving from their union with the Whole, thus deciding to de*part* from it. The famous role played by Charles Chaplin in his "Modern Times" is one good visual exemplification of this centrifugal factor. Gentle Giant's song "Cogs in Cogs" (Minnear, Shulman, & Shulman, 1974) is another eloquent example of this principle[5].

In the face of Parts-Whole mismatches, two are the viable options:
1. Enrolling Parts with a lesser systemic class. Technology sustains this option by providing specialized cyber-physical systems and robots that may be charged with the most repetitive and uncreative of tasks. A good example is give by domestic robots, such vacuum domotics or lawn motor domotics, that take charge of some of the household chores (DesMarais, 2013). Demoting the Parts to a systemic class closer to that of the Whole thus solves or softens the mismatch.

---

4 Sometimes expressed as the "affordances" and "constraints" built into a system or technology (Macdougall, 2013; 2014)

5 From (Minnear, Shulman, & Shulman, 1974): "Empty promise broken the path has / Not been paved any / Way. / Cogs in cogs the machine / Is being left where it lay. / Anger and the rising murmur breaks / The old circle, the wheel slowly turns around. / All words saying nothing / The air is sour with discontent. / No returns have been tasted / Or are they ever sent."

2.  Enhancing the systemic class of the Whole. This may be reached by improving the organizational structure of the Whole. Promoting the Whole thus solves or reduces the mismatch. An example of this approach can be found in the domain of Ambient Assisted Living (Sun et al., 2010): Moving from a strict hierarchy to a community of peer levels empowers all participants and enhances performance, e-Inclusion, and quality of life (De Florio and Blondia, 2010).

Let us consider the second strategy in a bit more detail. Two examples of organizational structures that enhance the systemic class of hierarchical systems should suffice.

# 4  Exception-based Hierarchies

As mentioned earlier, one strategy to enhance QoE is a reduction in the extent of Parts-Whole mismatches. This may be achieved by enhancing centripetal forces, for instance, by making it possible for Parts to exert organizational behaviors as close as possible to those typical of their own systemic class.

Crisis management provides a good example of this principle. Indeed a major flaw in the response to Hurricane Katrina was given by Parts-Whole mismatches: Institutional responders neglected the opportunity to "effectively use, collaborate with, and coordinate the combined public and private efforts" (Colten, Kates, & Laska, 2008), while it is now widely recognized how "empowering the citizens" and allowing them to play the role of informal responders effectively enhances the quality of the response (De Florio, Sun, & Blondia, 2014). Moving from the paradigm of "**remote control of the few**" to "**distributed control of the many**" results in greater QoE.

In order to enable these extended behaviors, the organization must *evolve*, for instance, from a pure hierarchy of enrolled Parts to what we call a hierarchy-with-exceptions: A hierarchy, that is, whose organizational rules allow occasional "violations" of the rigid separation into layers; and one in which the Parts are not passive constituents but rather active "Participants" (Sun et al., 2007) organized into semi-autonomous communities.

In what follows, the elements of two organizations representative of that model are introduced.

## 4.1  Sociocracy

A common way to overcome many of the limitations and risks associated with (human-based) hierarchical organizations is the sociocracy (Buck & Endenburg, 2012). Sociocracy introduces two simple rules that allow members in a layer to be temporarily promoted as members of the next layer up in the hierarchy. This happens through so-called sociocratic rules: Members of a layer, say layer $i$, meet regularly sharing their viewpoints and opinions about significant events; for instance, events pertaining to changes in the organization or the manifestation of unprecedented environmental conditions. Participating members are said to constitute a "circle". As a result of those meetings a member of the circle may be elected as representative of the circle. Thus, the elected representative becomes the new personification of a self-managed Whole (with the circle members being its Parts). This Whole is a new transient organizational entity that self-develops autonomically within the "greater Whole" of the organization.

The elected representative has a special, "double" nature in sociocracy:

- It is a member of layer $i$ of the hierarchy.

- At the same time, as representative of the circle of layer $i$, he or she is also a member of layer $i$+1 — albeit only temporarily.

Through its meetings and the preceding rule, sociocracy introduces temporary *exceptions* to the hierarchy. Through such exceptions the transient "Whole" may propagate control (taking the shape of, e.g., information, knowledge, analytical and planning insight) to the Parts in the next layer in the hierarchy. By means of its exceptions, in sociocracy the "**remote control of the few**" expands into a "**remote control of the few and the elected**". Experimentation in several contexts has proven that sociocracy, at least in certain cases, successfully enhances the resilience of human organizations throughout crises and turbulent conditions (for instance, economic competition) (Buck & Endenburg, 2012). We conjecture that this greater quality may be due to sociocracy's ability to reduce the extent of Parts-Whole mismatches.

A few observations are now of use:

- In sociocracy, a transient Whole only manifests itself between neighboring layers.

- The lifecycle of a transient Whole takes place within the boundaries of the same organization.

- The definition of a new transient Whole calls for human participation (through the circle meetings).

- (As a corollary to previous observation): Sociocracy only applies to human-based organizations.

Because of the above limitations, sociocracy may not be as easily applied to cases in which the organization regularly faces turbulent environmental conditions. Such conditions tend to require rapid establishment of new transient Wholes made of parts residing in multiple layers of the organization, or even of different organizations. As discussed, e.g., in (Colten, Kates, & Laska, 2008, De Florio et al., 2014), the emergence of "Community Wholes" is expected to significantly enhance an organization's ability to respond to crises and disastrous conditions (Community

Resilience) and to learn how to dynamically improve its community-environment fit (an ability that could be referred to as "Community Antifragility").

Sociocracy is usually applied to pre-existing organizations as an "ancillary" organizational structure. In fact, it does not propose a radical change in the control paradigms of the hierarchy: Control is still firmly in the hand of the entity personifying the hierarchy; only, "*control bubbles*" temporarily surface between neighboring layers. As a consequence of this phenomenon, the hierarchy becomes more "fluid" and in particular chances of failures and congestion are reduced while the agility and competitiveness are increased (Buck & Endenburg, 2012).

An organization that derived from sociocracy is holacracy (Robertson, 2007). Holacracy is also based on the sociocratic "axioms" of circle, meeting, and elected representative. A major difference of holacracy with respect to sociocracy lies in the fact that holacracy is explicitly based on a fractal organization, namely the recursive application of a set of rules and of an organizational structure that is "simultaneously a part and a whole, a container and a contained, a controller and a controlled" (Sousa, Silva, Heikkila, Kallingbaum, & Valcknears, 2000). Another difference is the fact that holacracy focuses on roles that may or may not be "fired" depending on the availability of actants.

## 4.2 Fractal Social Organizations

Fractal Social Organizations (FSOs) are a biologically inspired fractal organization that extends sociocracy. In what follows the focus will be only on those aspects of FSOs that best match the themes of the current volume. Readers are referred to (De Florio, Sun, Buys, & Blondia, 2013; De Florio, Bakhouya, Coronato, & Di Marzo Serugendo, 2013) for more details.

FSOs extend sociocracy with rules enabling the creation and the life-cycle management of "Community Wholes" not dissimilar from those mentioned in Sect. 4.1. In what follows, this is shown by describing how FSOs allow for the spontaneous and self-managed emergence of so-called "social overlay networks" (SONs), namely inter-layer and inter-organizational transient Wholes.

As in sociocracy, this is reached in FSOs through an exception mechanism. Contrarily to sociocracy, exception is not a result of electing a representative during circle meetings. FSOs introduce the concepts of firing conditions and role exceptions (De Florio et al., 2013): Conditions (such as an alarming situation) are autonomously detected in a region of the organization; whenever this takes place, a matching treatment protocol is selected (for instance, a fire rescue protocol). Protocols call for *roles* to be assigned to agents (e.g., a fire rescue squad; a firefighter truck; extinguishers; and so on). Assignment of roles is first attempted in the originating region. If one or more roles cannot be assigned, this triggers so-called "role exceptions" that are propagated throughout the hierarchy. If roles can be assigned in other layers, this produces a new intra-organizational transient Whole: the SON; otherwise, either the protocol fails or the exception is forwarded to a neighboring organization (if any can be located). Neighboring organizations do not necessarily take the shape of another full-fledged FSO; in fact any complying entity willing to play the missing role — including for instance a team of citizens acting as informal responders in a community resilience scenario (see Sect. 4) — would become parts of the new inter-organizational Whole. This matches the requirement expressed in (Colten et al., 2008) and referred to as "a central task of enhancing community resilience," namely the ability to "effectively use, collaborate with, and coordinate the combined public and private efforts [...] in advance of hazard events".

It is worth remarking what follows:

- In any organization, Parts-Whole mismatches occur during the enrollment process. When several options are available, the FSOs can choose to select as SON members those actors that minimize Parts-Whole mismatches.
- Exceptions are not managed in FSOs through circle meetings and elections of representatives; rather, an exception triggers the execution of a role assignment protocol. Semantic service description and matching makes it possible to execute those protocols in a fully automated way as described, e.g., in (De Florio, Sun, & Bakhouya, 2014, Sun, De Florio, & Blondia, 2013, De Florio & Blondia, 2010).
- Inter-organizational cooperation rules may be defined, making it possible for instance, for a defense organization to "lend" actors to an emergency management body — provided that an inter-organizational semantic service description and matching is in use.

Thus, FSOs can be structured in such a way as to overcome all limitations of sociocracy identified in Sect. 4.1.

A particular aspect of QoE, namely controllability, requires particular attention in FSOs. The persona emerging from an FSO is in fact much less determined and stable than in sociocracy. Due to the dynamic nature of Parts, which can be controller and controlled entities depending on contingent causes, an FSO may "host" multiple personae. As a consequence, no single "remote control" thread exists in FSOs; rather, multiple "control bubbles" may be concurrently "floating" within one FSO. If precautions are not taken, such control bubbles may even not cooperate towards the same end. Control is thus distributed and highly dynamic, which makes verification more difficult. Coupling an FSOs with a traditional organization and confining the scope of the services supplied by the FSOs is a strategy suggested in (De Florio et al., 2013) to deal with this problem.

# 5 Conclusions

This discussion has considered the problem of the quality of emergence in the organization of a set of interacting constituents. We observed how this quality may not be considered as a static and immutable "fact"; rather, it should be recognized as the result of dynamic interactions between forces of opposite value. By recalling the System of Leibniz, two major classes of factors determining quality of emergence were identified. Two organizational structures — traditional hierarchies and exception-based hierarchies — were then described. It was observed how, in both cases, organizational *axioms* influence the quality of emergence in different ways. Two classes of exception-based hierarchies were then discussed: Sociocracy and Fractal Social Organizations. We considered how, at least in certain cases, the addition of simple "cybernetic rules" to an organization resulted in a significant improvement in an organization's ability to withstand turbulence and competition, as well as in its agility and competitiveness. Finally we observed how Fractal Social Organizations make it possible to compose complex intra-organizational "social overlay networks". These "organizations within organizations" are temporary networks of organs whose constituents may be enrolled from any layer of the greater organization and whose purpose is to bring about the best "organizational response" to a context change. Examples here might include harsher environmental conditions, impending threats, or unprecedented opportunities.

A lesson learned though the present discussion pertains to the main theme of this volume. Moving from the paradigm of "remote control of the one", or "the few", to that of the "distributed control of the many" resulted in conflicting results. Greater QoE may be observed in certain cases, although in other cases enhanced agility and adaptability may jeopardize controllability and determinism. Leaving the "safe shores" of the McLuhanian vision of organizations as technological extensions of an individual Whole promises unprecedented degrees of organizational efficiency; at the same time, novel mechanisms are becoming necessary in order to guarantee persistence of identity and, in the long run, avoid scenarios such as those recently prophesied by Stephen Hawking (2014):

> "Whereas the short-term impact of AI depends on who controls it, the long-term impact depends on whether it can be controlled at all."

Further reflections on this final theme can be found in (De Florio, 2013).

## Acknowledgments

I would like to express my gratitude to the Editor of the present volume, Dr. Robert MacDougall, for his support, encouragement, and very valuable suggestions.